# Momentum-space theory for topological magnons in 2D ferromagnetic skyrmion lattices

Doried Ghader*[1] and Bilal Jabakhanji[1]

[1] College of Engineering and Technology, American University of the Middle East, Egaila 54200, Kuwait

*doried.ghader@aum.edu.kw

## Abstract

Magnon dynamics in skyrmion lattices have garnered significant interest due to their potential applications in topological magnonics. Existing theories often follow a single-momentum approach, assuming significant Dzyaloshinskii-Moriya Interaction (DMI) to minimize the skyrmion's dimensions, which can lead to oversimplification in describing magnon behavior. This study introduces a multi-momentum operator theory for magnons in realistically large 2D skyrmions, where each skyrmion encompasses several thousand spins. The proposed theory fully transforms the magnon Hamiltonian into momentum space, incorporating off-diagonal terms to capture umklapp scattering caused by the skyrmion wave vectors. Our results reveal deviations from single-momentum theories, demonstrating that flat bands are not universal features of the skyrmionic magnon spectrum. Additionally, we find that manipulating the skyrmion size with an external magnetic field induces multiple topological phase transitions. At high magnetic fields, the low-energy magnon spectrum becomes densely packed and entirely topological, resembling a topological band continuum.

## I. Introduction

Skyrmion lattices are compelling platforms for probing unconventional magnon dynamics [1–5], holding significant potential for advancements in topological magnonics [6]. Theories [1,7–12] suggest that the topological winding of the skyrmion generates a fictitious magnetic field that acts upon magnons, thereby influencing their dynamical behavior. This emergent interaction between magnons and the underlying spin texture leads to topological magnon bands with finite Berry curvatures and nonzero Chern numbers. Recent experiments [4,13] supported the theoretical predictions, presenting solid evidence of topological magnon bands and thermal magnon Hall effect in skyrmion lattices.

Operator-based theories on skyrmionic magnons [3,7–9,14–16] commonly treat the skyrmion lattice as a cluster of real-space sublattices, where each spin in the unit cell (representing a single skyrmion) introduces a real-space sublattice magnon operator. In this framework, a large Dzyaloshinskii-Moriya Interaction (DMI) is often assumed to minimize the dimensions of the skyrmion. By maintaining a real-space cluster description, the resulting Hamiltonian becomes a single-momentum operator that forces magnons to travel with uniform momenta, potentially oversimplifying the complex dynamics of magnons in skyrmion lattices.

In most materials, the DMI and magnetic fields stabilizing the skyrmion lattice are typically weak [17], resulting in skyrmions that encompass several thousand spins. Insights into the inverse relationship between DMI strength and skyrmion size in a skyrmion lattice can be found in Figure 1 and references [18,19]. For large skyrmions, the real-space cluster approach becomes less practical for studying magnons and may be insufficient to fully capture the effects of the spin texture on magnon dynamics. These skyrmions effectively form a continuum that can be reconstructed via a set of characteristic wave vectors. The skyrmion wave vectors are pivotal in shaping the magnon dynamics, as they lead to umklapp scattering events that alter the magnon's momentum. Capturing this complex interaction landscape and its effect on magnon dynamics is beyond the reach of single-momentum Hamiltonians, necessitating a theory that encodes the skyrmion wave vectors into the magnons' operators.

This work introduces a momentum-space continuum theory to explore magnon dynamics in large 2D ferromagnetic skyrmions. Employing Fourier-space techniques, we construct a momentum-space representation of the skyrmion lattice, associating magnon operators with sites defined by the skyrmion's wave vectors. Utilizing these operators, we derive a multi-momentum magnon Hamiltonian, with diagonal and off-diagonal terms capturing ordinary and umklapp magnon scatterings, respectively. The magnon spectra obtained from this approach underscore the crucial role of skyrmion size, determined by the magnetic field for a given DMI, in shaping magnon dynamics. The magnetic field compacts the magnon spectrum, leaving small gaps between adjacent bands that enable valid definitions of the Berry curvatures and Chern numbers. The number of topological bands increases with higher magnetic fields, while flat bands are rarely observed, even at a minimal magnetic field. This suggests that the magnetic field induces multiple topological phase transitions, rather than a single transition at a critical magnetic field value [9]. Notably, flat bands are absent at the maximal magnetic field, and the entire low-energy magnon spectrum becomes topological and densely packed, resembling a continuum of topological bands.

## II. Theory

This section concisely presents our theory, while detailed derivations are provided in the Supplementary Notes [20]. We start by considering a 2D triangular lattice of spins on the xy-plane, described by the real-space Hamiltonian:

$$\mathcal{H} = -J\sum_{i,j} \boldsymbol{\mathcal{S}}_i \cdot \boldsymbol{\mathcal{S}}_j - \sum_{i,j} \boldsymbol{D}_{ij} \cdot \boldsymbol{\mathcal{S}}_i \times \boldsymbol{\mathcal{S}}_j - B\sum_{i} \mathcal{S}_{\boldsymbol{i}}^{\boldsymbol{z}} \tag{1}$$

In Equation 1, $\boldsymbol{\mathcal{S}}_i$ denotes the spin operator at site $i$. The terms in order represent the nearest neighbor (NN) ferromagnetic exchange, chiral NN DMI [21,22] compatible with interfacial inversion symmetry breaking [17,23], and Zeeman coupling due to an external magnetic ($\boldsymbol{B}$) along the $z$-axis. The DMI vectors are given by $\boldsymbol{D}_{ij} = D\, \hat{\boldsymbol{z}} \times \hat{\boldsymbol{r}}_{ij}$, where $\hat{\boldsymbol{z}}$ and $\hat{\boldsymbol{r}}_{ij}$ are unit vectors along the z-axis and the NN bond, respectively. The coefficients $J$ and $D$ quantify the strengths of the exchange and DMI, respectively.

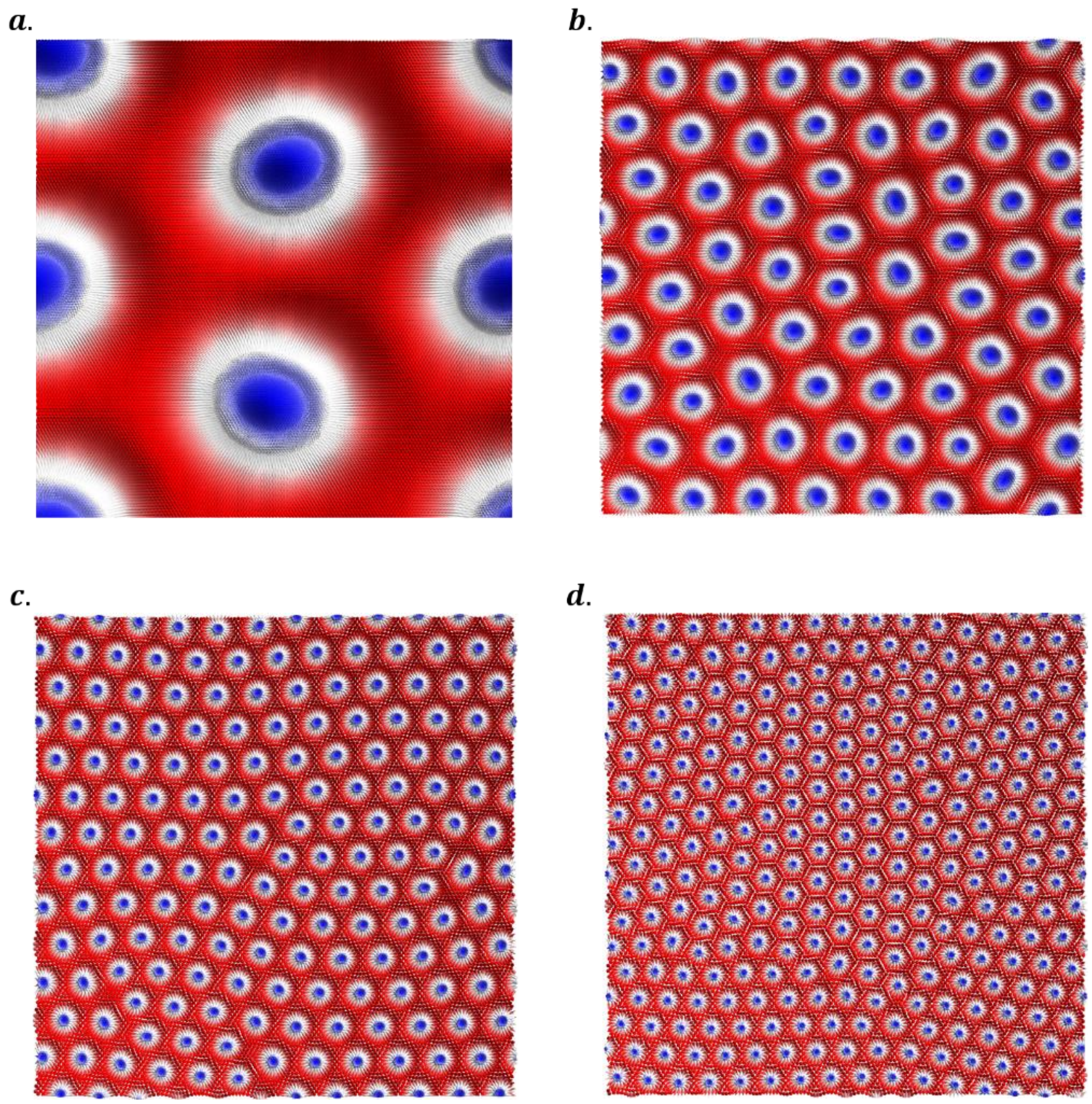

**Figure 1.** Skyrmion lattices obtained from stochastic Landau-Lifshitz-Gilbert simulations with DMI strengths of $D = 0.1\,J$ (a), $D = 0.4\,J$ (b), $D = 0.7\,J$ (c), and $D = 1\,J$ (d). The corresponding magnetic fields stabilizing these skyrmion lattices are $B = 0.13\,T$ (a), $B = 1.8\,T$ (b), $B = 5.8\,T$ (c), and $B = 10.5\,T$ (d). The simulations were performed on samples with $24000$ spins, with the magnetic field applied along the $+z$ direction. The temperature-color map represents the z-component of the normalized spin, ranging from $-1$ (blue) to 1 (red).

For suitable values of $B$, the classical ground state of the Hamiltonian forms a ferromagnetic skyrmion lattice [3,4,7–9,24–29], which we simulate using the stochastic Landau-Lifshitz-Gilbert equations (sLLG) within the Vampire software [30]. Figure 1 presents our results for different values of the DMI in a system comprising $24000$ spins. The simulations are initiated from random spin configurations at high temperatures and progressively cooled to near $0\,K$. The figure illustrates the decrease in skyrmion size with increasing DMI strength. Hereafter, we will focus on large skyrmions formed at $D = 0.1\,J$. For this specific $D$ value, multiple simulations with gradually increasing magnetic field strengths revealed the formation of Néel-type skyrmions within the magnetic field range of $0.12\,T \lesssim B \lesssim 0.225\,T$. However, due to the random nucleation of DMI-induced skyrmions [31–34], achieving perfectly ordered skyrmion lattices proved challenging (Figure 1). Therefore, we employed suitable functions to model the skyrmion lattice precisely, after determining the unit cell size from the Vampire simulations (Supplementary Notes [20]).

The real-space magnon Hamiltonian can be derived from Equation 1 by employing standard rotation techniques [7–9,35]. Specifically, Rodrigues' rotation formula is utilized to define a rotated frame, aligning its z-axis with the local spin direction within the skyrmionic ground state. Therefore, the classical ground state manifests as a ferromagnetic configuration in the rotated frame. This allows for the representation of rotated spin operators in terms of magnon creation ($a_i^+$) and annihilation ($a_i$) operators through the Holstein-Primakoff transformation [36]. By incorporating the rotated spin operators and their Holstein-Primakoff representations into Equation 1 (the detailed derivation is provided in the Supplementary Notes [20]), we determine the real-space magnon Hamiltonian as follows,

$$\mathcal{H} = \mathcal{H}^{ex} + \mathcal{H}^{D} + \mathcal{H}^{Zee} \tag{2a}$$

with,

$$\mathcal{H}^{ex} = JS \sum_{i,j,m} \left[\tfrac{1}{2} R_{mz,i}\, R_{mz,j}\, a_i^+ a_i - R_{m-,i}\, R_{m+,j}\, a_i^+\, a_j - R_{m-,i}\, R_{m-,j}\, a_i^+\, a_j^+\right] + h.c. \tag{2b}$$

$$\mathcal{H}^{D} = S \sum_{i,j} \sum_{m,m_1,m_2} \in_{m m_1 m_2}\, D_j^m \left[\tfrac{1}{2} R_{m_1 z,i}\, R_{m_2 z,j}\, a_i^+ a_i - R_{m_1 -,i}\, R_{m_2 +,j}\, a_i^+\, a_j - R_{m_1 -,i}\, R_{m_2 -,j}\, a_i^+\, a_j^+\right] + h.c. \tag{2c}$$

$$\mathcal{H}^{Zee} = BS\sum_{i} R_{zz,i} a_i^{+} a_i + h.c.$$

(2d)

In Equation 2, the Hamiltonians $\mathcal{H}^{ex}$, $\mathcal{H}^{D}$, and $\mathcal{H}^{Zee}$ account for the exchange, DMI, and Zeeman contributions, respectively. The term '$h.c.$' denotes Hermitian conjugation. The indices $m$, $m_1$, and $m_2$ are summed over $x$, $y$, and $z$. $D_j^m$ and $\in_{mm_1m_2}$ stand for the components of the DMI vectors and the Levi-Civita symbol, respectively.

We aim to transform $\mathcal{H}$ from the real space to the Fourier (or momentum) space, thereby removing its dependency on the real space sites ($i$ and $j$ in Equation 2). This transition entails Fourier expansions of the rotation matrix elements $R_{mn}$, utilizing characteristic wave vectors defined by the skyrmion lattice. The optimal set of skyrmion wave vectors is derived from the Fourier expansion of the classical spin vectors that form the skyrmion lattice ground state.

As depicted in Figure 2b, the skyrmions form a triangular Bravais lattice with large real-space lattice vectors $\boldsymbol{a}_1$ and $\boldsymbol{a}_2$, and a correspondingly tiny Brillouin Zone (BZ), defined by reciprocal lattice vectors $\boldsymbol{b}_1$ and $\boldsymbol{b}_2$. The classical spin vector varies over a length scale much larger than the lattice spacings $a$, necessitating a continuum approach. In this approach, the classical ground state is represented as a continuous vector field $\boldsymbol{\mathcal{S}}(\boldsymbol{r})$, which admits a Fourier series expansion in the form

$$\boldsymbol{\mathcal{S}}(\boldsymbol{r}) = \sum_{\gamma,\sigma\epsilon h} \left[\boldsymbol{\mathcal{S}}^{\gamma\sigma} e^{i\mathbf{G}_{\gamma,\sigma}\cdot\boldsymbol{r}} + \boldsymbol{\mathcal{S}}^{\gamma\sigma *} e^{-i\mathbf{G}_{\gamma,\sigma}\cdot\boldsymbol{r}}\right]$$

(3)

Here, $\mathbf{G}_{\gamma,\sigma} = \gamma\boldsymbol{b}_1 + \sigma\boldsymbol{b}_2$ ($\gamma$ and $\sigma$ are integers) represent the skyrmion wave vectors. The functional form of the continuous field $\boldsymbol{\mathcal{S}}(\boldsymbol{r})$ is presented in the Supplementary Notes [20]. The vector coefficients $\boldsymbol{\mathcal{S}}^{\gamma\sigma}$ are calculated using the Fourier transform formula, $\frac{1}{A}\iint d^2r\, \boldsymbol{\mathcal{S}}(\boldsymbol{r}) e^{\pm i\mathbf{G}_{\gamma,\sigma}\cdot\boldsymbol{r}}$, performed numerically over the unit cell (a single skyrmion) with area $A$.

Our numerical analysis demonstrates that the Fourier series in Equation 3 yields excellent convergence with 64 wave vectors $\mathbf{G}_{\gamma,\sigma}$, compiled in the set $h = \{(0,\sigma) \mid 0 \le \sigma \le 6\} \cup \{(\gamma,\sigma) \mid 1 \le \gamma \le 6\ and\ \gamma - 6 \le \sigma \le 6\}$. The vectors $\boldsymbol{\mathcal{S}}^{\gamma\sigma}$ can be expressed as $\boldsymbol{\mathcal{S}}^{\gamma\sigma} = \langle i\mathcal{S}_x^{\gamma\sigma}, i\mathcal{S}_y^{\gamma\sigma}, \mathcal{S}_z^{\gamma\sigma}\rangle$, outlining a momentum-space spin texture (Figure 2c) comprising spins $\langle \pm\mathcal{S}_x^{\gamma\sigma}, \pm\mathcal{S}_y^{\gamma\sigma}, \mathcal{S}_z^{\gamma\sigma}\rangle$ located at sites $\pm\mathbf{G}_{\gamma,\sigma}$, respectively. As expected, the magnitudes of the momentum-space spins $\boldsymbol{\mathcal{S}}^{\gamma\sigma}$ diminish significantly beyond the second BZ, as demonstrated in

Figure 2d. This behavior is crucial for ensuring the convergence of the Fourier series expansion in Equation 3.

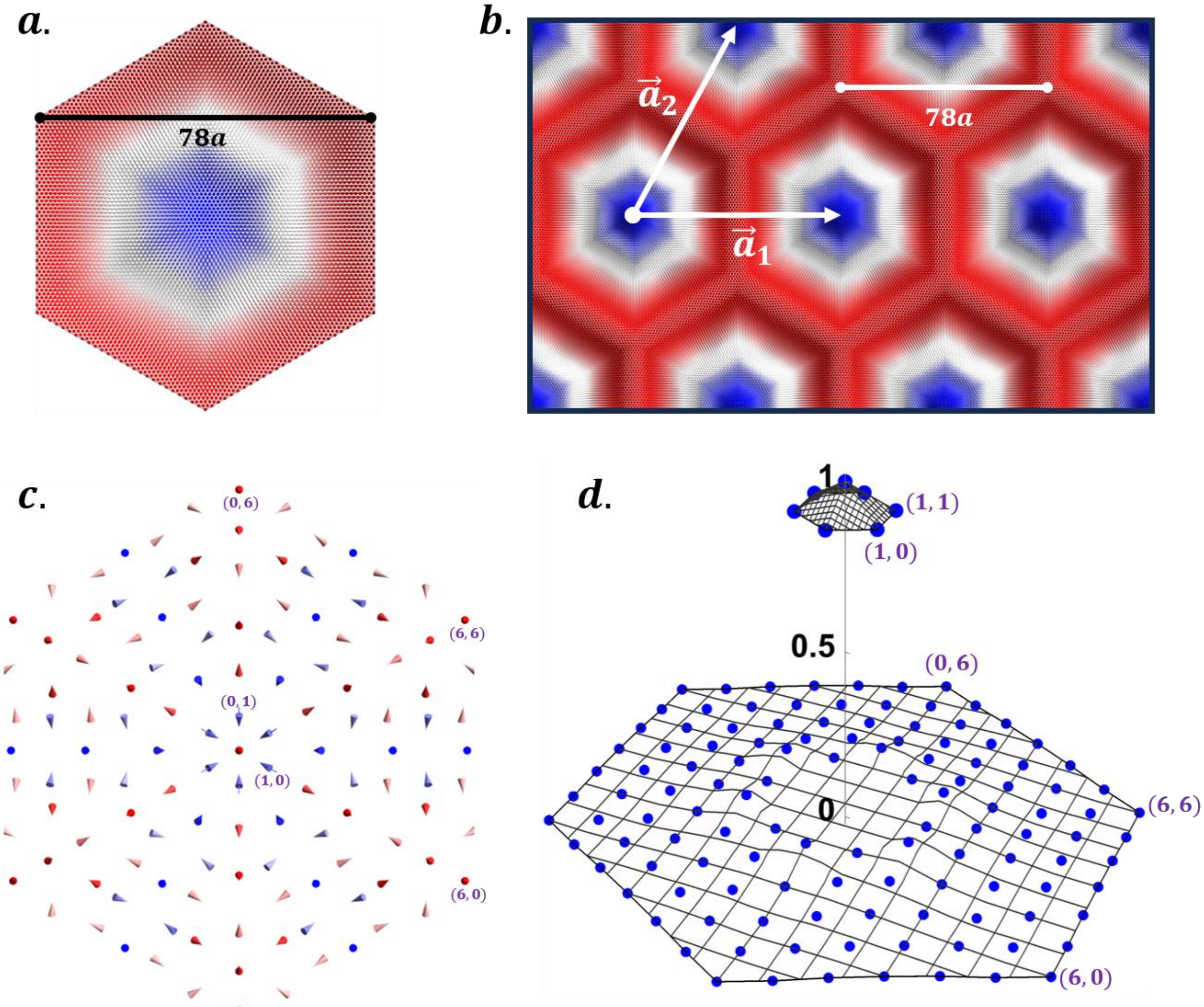

**Figure 2.** (a) A single Néel-type skyrmion with a DMI strength of $D = 0.1\,J$ and a magnetic field of $B = 0.12\,T$, encompassing 6163 spins. The skyrmion's width is $78a$, where $a$ represents the lattice spacing between the spins. The temperature-color map represents the z-component of the normalized spin, ranging from $-1$ (blue) to 1 (red). The magnetic field is applied along $+z$. (b) The skyrmion lattice under these parameters, with $\boldsymbol{a}_1$ and $\boldsymbol{a}_2$ representing the real-space lattice vectors. The lattice period corresponds to the skyrmion width ($78a$). (c) The momentum-space representation of the skyrmion lattice, with momentum-space spins located at positions $(\gamma, \sigma)$, defined by the skyrmion wave vectors $\mathbf{G}_{\gamma,\sigma} = \gamma \boldsymbol{b}_1 + \sigma \boldsymbol{b}_2$, where $\boldsymbol{b}_1$ and $\boldsymbol{b}_2$ are the reciprocal-space lattice vectors. (d) The magnitudes of the momentum-space spins normalized with respect to the spin at the BZ center.

Next, we utilize Rodrigues' rotation formula to express the rotation matrix elements $R_{mn}$ in terms of the components of the continuous field $\boldsymbol{S}(\boldsymbol{r})$, as detailed in the Supplementary Notes [20]. Consequently, these elements are defined as continuous functions $R_{mn}(\boldsymbol{r})$ and are expanded into Fourier series using the same set $h$ of skyrmion wave vectors,

$$R_{mn}(\boldsymbol{r}) = \sum_{\gamma,\sigma \epsilon h} \left[\bar{Q}_{mn}^{\gamma\sigma} e^{i\mathbf{G}_{\gamma,\sigma}\cdot\boldsymbol{r}} + Q_{mn}^{\gamma\sigma} e^{-i\mathbf{G}_{\gamma,\sigma}\cdot\boldsymbol{r}}\right] \tag{4}$$

The Fourier coefficients $Q_{mn}^{\gamma\sigma}$ and $\bar{Q}_{mn}^{\gamma\sigma}$, with $m \in \{x, y, z\}$ and $n \in \{+, -, z\}$, are computed numerically using the Fourier transform formula $\frac{1}{A}\iint d^2r\, R_{mn}(\boldsymbol{r}) e^{\pm i\mathbf{G}_{\gamma,\sigma}\cdot\boldsymbol{r}}$.

In turn, the magnon operators are expressed in the standard Fourier form, $a_i^+ = \frac{1}{\sqrt{N}}\sum_{\boldsymbol{k}} e^{-i\boldsymbol{k}\cdot\boldsymbol{r}_i} a_{\boldsymbol{k}}^+$, where $N$ represents the total number of lattice sites, and $\boldsymbol{k}$ is a momentum within the first BZ. By substituting the Fourier expansions of $R_{mn}(\boldsymbol{r})$ into $\mathcal{H}$ (Equation 2) and executing the summations over the real-space lattice sites, we successfully derive the momentum-space magnon Hamiltonian. The derivation is lengthy, and the details are presented in the Supplementary Notes [20]. The derivation demonstrates how the coupling between $\boldsymbol{k}$ and the skyrmion lattice wave vectors $\boldsymbol{G}_{\gamma,\sigma}$ introduces the multi-momentum bosonic operators $a_{\boldsymbol{k}\pm\boldsymbol{G}_{\gamma,\sigma}}^+$ and $a_{\boldsymbol{k}\pm\boldsymbol{G}_{\gamma,\sigma}}$ into the magnon Hamiltonian. The emergence of these operators is a manifestation of the umklapp scattering processes experienced by magnons while traveling through the skyrmion lattice.

The resulting momentum-space Hamiltonian can be written in compact form as follows,

$$\mathcal{H} = \frac{1}{2}\sum_{\boldsymbol{k}}\sum_{\gamma,\sigma\epsilon h}\left[\chi_{\gamma,\sigma;I} a_{\boldsymbol{k}}^+ a_{\boldsymbol{k}+\mathbf{G}_{\gamma,\sigma}} + \tilde{\chi}_{\gamma,\sigma;I} a_{-\boldsymbol{k}}^+ a_{-\boldsymbol{k}-\boldsymbol{G}_{\gamma,\sigma}} + \chi_{\gamma,\sigma;II} a_{\boldsymbol{k}}^+\, a_{-\boldsymbol{k}-\boldsymbol{G}_{\gamma,\sigma}}^+ + \tilde{\chi}_{\gamma,\sigma;II} a_{\boldsymbol{k}+\boldsymbol{G}_{\gamma,\sigma}}^+ a_{-\boldsymbol{k}}^+\right] + h.c. \tag{5}$$

The coefficients $\chi_{\gamma,\sigma;I}$, $\tilde{\chi}_{\gamma,\sigma;I}$, $\chi_{\gamma,\sigma;II}$, and $\tilde{\chi}_{\gamma,\sigma;II}$ are lengthy functions of $\boldsymbol{k}$ and $\boldsymbol{G}_{\gamma,\sigma}$. Their explicit expressions are provided in the Supplementary Notes [20].

To further explore the structure of $\mathcal{H}$ and its physical implications, we introduce the operator wave function $\Psi_{\boldsymbol{k}}^\dagger = [\phi_{\boldsymbol{k}}^\dagger \quad \phi_{-\boldsymbol{k}}^\dagger]$, with $\phi_{\boldsymbol{k}}^\dagger = \left[a_{\boldsymbol{k}+\boldsymbol{G}_{\gamma_1,\sigma_1}}^+ \quad \cdots \quad a_{\boldsymbol{k}+\boldsymbol{G}_{\gamma_n,\sigma_n}}^+ \quad a_{\boldsymbol{k}-\boldsymbol{G}_{\gamma_2,\sigma_2}}^+ \quad \cdots \quad a_{\boldsymbol{k}-\boldsymbol{G}_{\gamma_n,\sigma_n}}^+\right]$, and $\phi_{-\boldsymbol{k}}^\dagger = \left[a_{-\boldsymbol{k}-\boldsymbol{G}_{\gamma_1,\sigma_1}}^+ \quad \cdots \quad a_{-\boldsymbol{k}-\boldsymbol{G}_{\gamma_n,\sigma_n}}^+ \quad a_{-\boldsymbol{k}+\boldsymbol{G}_{\gamma_2,\sigma_2}}^+ \quad \cdots \quad a_{-\boldsymbol{k}+\boldsymbol{G}_{\gamma_n,\sigma_n}}^+\right]$. This allows Equation 5 to be recast in a matrix form as,

$$\mathcal{H} = \frac{1}{2}\sum_{\boldsymbol{k}} \Psi_{\boldsymbol{k}}^{\dagger}\mathcal{H}\left(\boldsymbol{k}, \boldsymbol{G}_{\gamma,\sigma}\right)\Psi_{\boldsymbol{k}} \tag{6}$$

with $\mathcal{H}\left(\boldsymbol{k}, \boldsymbol{G}_{\gamma,\sigma}\right) = \begin{bmatrix} H\left(\boldsymbol{k}, \boldsymbol{G}_{\gamma,\sigma}\right) & \Delta\left(\boldsymbol{k}, \boldsymbol{G}_{\gamma,\sigma}\right) \\ \Delta^{\dagger}\left(\boldsymbol{k}, \boldsymbol{G}_{\gamma,\sigma}\right) & \widetilde{H}\left(\boldsymbol{k}, \boldsymbol{G}_{\gamma,\sigma}\right) \end{bmatrix}$.

The block matrices within $\mathcal{H}\left(\boldsymbol{k}, \boldsymbol{G}_{\gamma,\sigma}\right)$ are of dimensions $n \times n$, where $n = 64$ represents the number of relevant sites in momentum space, as specified by the set $h$. These blocks elucidate the permitted magnon umklapp scatterings in momentum space. For instance, an umklapp scattering event from $\boldsymbol{k} + \boldsymbol{G}_{\gamma_i,\sigma_i}$ to $\boldsymbol{k} + \boldsymbol{G}_{\gamma_j,\sigma_j}$ occurs if and only if there exists a pair $(\gamma_l, \sigma_l) \in h$ such that $(\gamma_i, \sigma_i) + (\gamma_l, \sigma_l) = \left(\gamma_j, \sigma_j\right)$. Such scattering generates the matrix elements $H_{ij} = \chi_{\gamma_l,\sigma_l;I}\left(\boldsymbol{k} + \boldsymbol{G}_{\gamma_i,\sigma_i}\right)$, $\Delta_{ij} = \chi_{\gamma_l,\sigma_l;II}\left(\boldsymbol{k} + \boldsymbol{G}_{\gamma_i,\sigma_i}\right)$, $\Delta_{ji} = \tilde{\chi}_{\gamma_l,\sigma_l;II}\left(\boldsymbol{k} + \boldsymbol{G}_{\gamma_i,\sigma_i}\right)$, and $\widetilde{H}_{ji} = \tilde{\chi}_{\gamma_l,\sigma_l;I}(\boldsymbol{k} + \boldsymbol{G}_{\gamma_i,\sigma_i})$. Conversely, the reverse umklapp scattering, $\left(\gamma_j, \sigma_j\right) - (\gamma_l, \sigma_l) = (\gamma_i, \sigma_i)$, leads to the Hermitian conjugate matrix elements. By systematically accounting for all allowed umklapp scattering processes, we construct the matrix $\mathcal{H}\left(\boldsymbol{k}, \boldsymbol{G}_{\gamma,\sigma}\right)$, which is inherently Hermitian. To determine the magnonic band structure, we apply the standard Bogoliubov transformation to diagonalize $\mathcal{H}\left(\boldsymbol{k}, \boldsymbol{G}_{\gamma,\sigma}\right)$, utilizing the numerical approach developed by Colpa [37].

Before concluding this section, we revisit the expression $\left(a_{\boldsymbol{k},1}, \dots, a_{\boldsymbol{k},N}, a^{+}_{-\boldsymbol{k},1}, \dots, a^{+}_{-\boldsymbol{k},N}\right)^{T}$ for the magnon wave function in the real-space cluster approach [3,7–9,14–16]. This wave function collects the magnon operators $a_{\boldsymbol{k},i}$ and $a^{+}_{-\boldsymbol{k},i}$, associated with the real-space sites $i = 1, \dots, N$ of the skyrmion, all sharing the same momentum $\boldsymbol{k}$. Using Fourier techniques applicable specifically to large skyrmions, we replaced these operators with multi-momentum operators that capture the umklapp scattering processes.

## III. Magnon bands and topology

For the DMI strength $D = 0.1\,J$ chosen in our study, the minimum magnetic field required to stabilize skyrmions is approximately $0.12\,T$, as determined through the sLLG simulations described in the previous section. The skyrmion reaches its maximum size under this minimal magnetic field, comprising 6163 spins (Figures 2a, b). Due to their Bosonic nature, magnons tend to accumulate in the low-energy bands, which are therefore the primary focus of our calculations. Figure 3a displays the lowest 15 magnonic bands of this skyrmion ground state, calculated along the symmetry axes depicted in Figure 3b. Henceforth, we label these bands sequentially from $n = 1$ to $n = 15$, in order of increasing energy.

Given the large size of the skyrmions and the correspondingly tiny BZ, the 15-band magnon spectrum is confined to low energies, not exceeding $0.2J$ (Figure 3a). The lowest-energy band ($n = 1$) remains nonzero at the Γ point (the BZ's center), indicating the absence of Goldstone modes. We observe a global bandgap of approximately $0.016J$ between the first and second bands, with the latter appearing notably flat compared to all other bands. Despite its apparent flatness, a 3D plot of this band reveals its dispersion within a narrow bandwidth, as depicted in Figure 3c. The second and third modes are separated by a sizeable global bandgap ($\sim 0.05J$), while a smaller gap of about $0.017J$ separates the fourth and fifth modes. The low-energy bands $n = 1, \dots, 4$ stand out as relatively isolated from the remaining bands, while the magnon spectrum becomes more compact beyond these bands. At higher energies, the bands exhibit several avoided crossing points where they approach each other without closing the gaps at this specific magnetic field. This phenomenon leads to the formation of numerous small bandgaps, examples of which are depicted in Figures 3d and 3e.

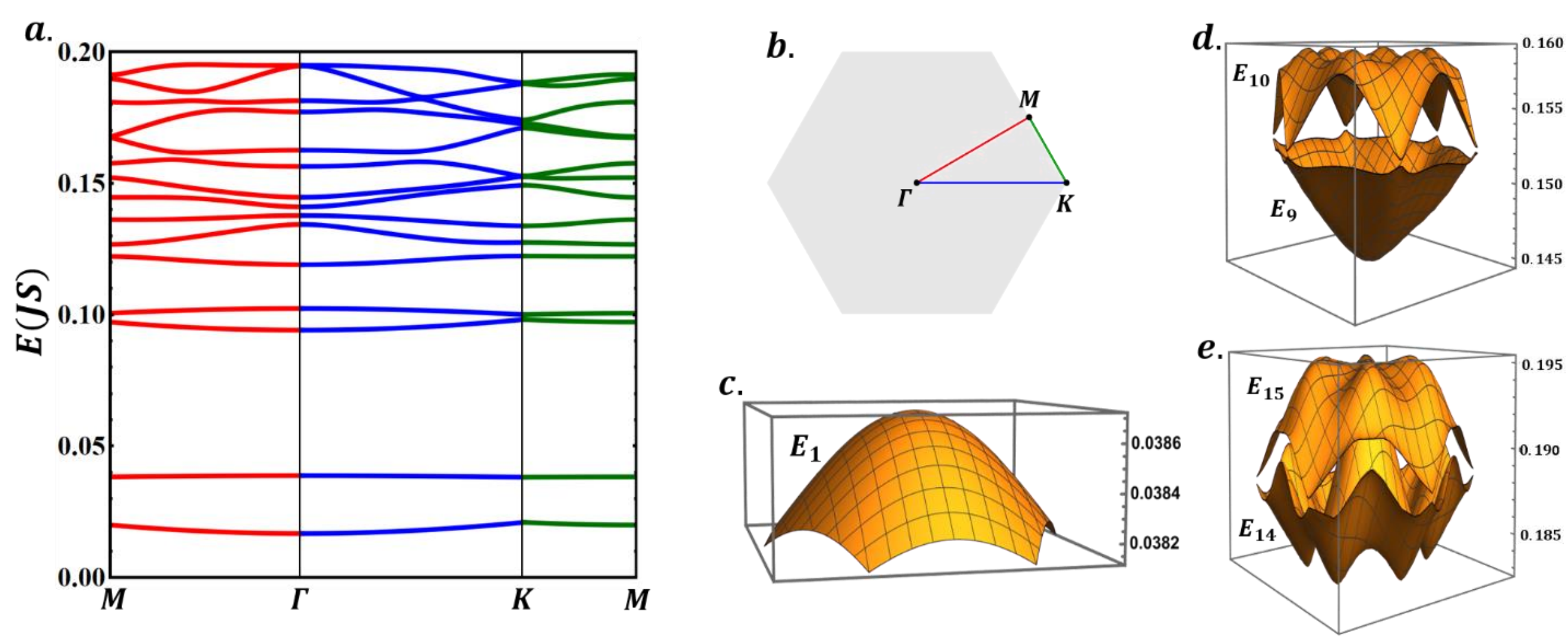

**Figure 3.** (a) Dispersion curves for the lowest 15 magnon bands of the skyrmion ground state at $D = 0.1J$ and $B = 0.12\ T$. The bands are plotted along the high-symmetry axes of the BZ shown in (b). (c) A 3D plot over the BZ of $E_2$ (the second energy band), revealing its dispersion within a narrow bandwidth. (d) A 3D plot over the BZ showing the tiny gaps between bands $E_9$ and $E_{10}$. (e) A similar plot to (d), but for bands $E_{14}$ and $E_{15}$.

The gaped magnonic spectrum motivates the analysis of its topological features. Using the numerical method developed by Fukui *et al.* [38], we calculated the bands' Berry curvatures and Chern numbers. Details on the application of this numerical approach to multi-band spectra can be found in the Supplementary Notes [20] and other references [39–41]. Our results for the Chern numbers ($C_n, n = 1, \dots, 15$) are summarized in Table 1, and the Berry curvatures for the topological bands are presented in Figure 4**.** However, under the minimal magnetic field, the ground state may not be optimal for observing topological effects, as the lowest-energy bands are found to be topologically trivial (Table 1).

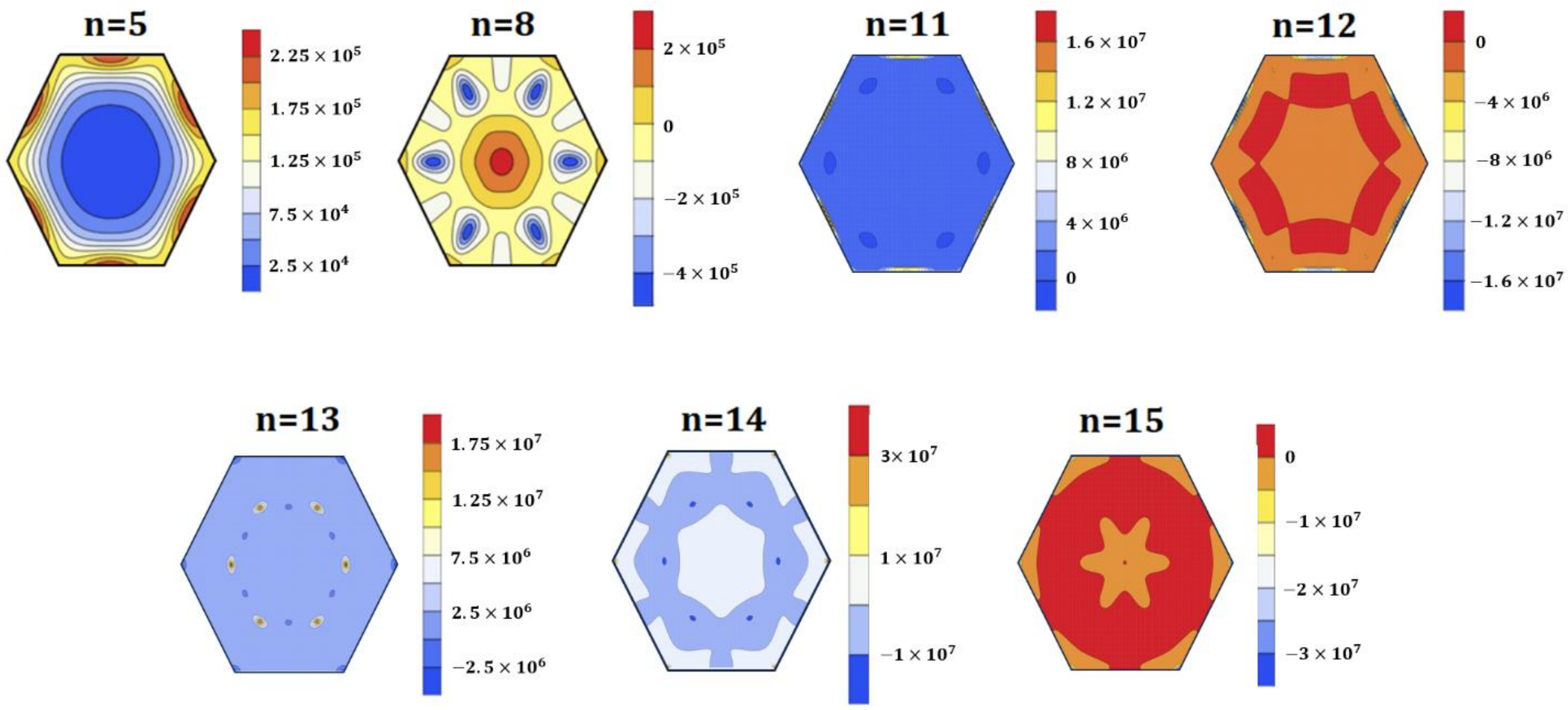

**Figure 4.** Berry curvatures for the topological magnon bands at a magnetic field of $0.12\ T$. The corresponding Chern numbers are listed in the second row of Table 1.

In our theory, the Hamiltonian $\mathcal{H}(\boldsymbol{k}, \boldsymbol{G}_{\gamma,\sigma})$ is critically dependent on the reciprocal lattice vectors $\boldsymbol{b}_1$ and $\boldsymbol{b}_2$, which are determined by the size of the skyrmion. Thus, within the magnetic field range that stabilizes the skyrmion lattice, $0.12\ T \lesssim B \lesssim 0.225\ T$ at $D = 0.1\ J$, an increase in the magnetic field results in a reduction of the skyrmion size, which significantly impacts the magnonic band structure. Specifically, at $B = 0.175\ T$, near the middle of the magnetic field range, the skyrmion size diminishes to 3661 spins, as illustrated in Figures 5a and 5b. The associated magnon band structure, shown in Figure 5c, differs markedly from that at the minimum magnetic field. The spectrum becomes more compact, reaching higher energies ($\sim 0.3\ J$) because of the larger reciprocal lattice vectors $\boldsymbol{b}_1$ and $\boldsymbol{b}_2$. Furthermore, the shift of the lowest energy mode to higher energies demonstrates the impact of skyrmion size—and, consequently, the magnetic field—on the energy gap of this mode. Although the lowest-energy band appears flat, closer examination shows that it is dispersive within a narrow range (Figure 5d). The two lowest-energy bands remain relatively separated from the rest, while the spectrum's high-energy region is notably compact, featuring numerous tiny bandgaps at points of avoided crossing.

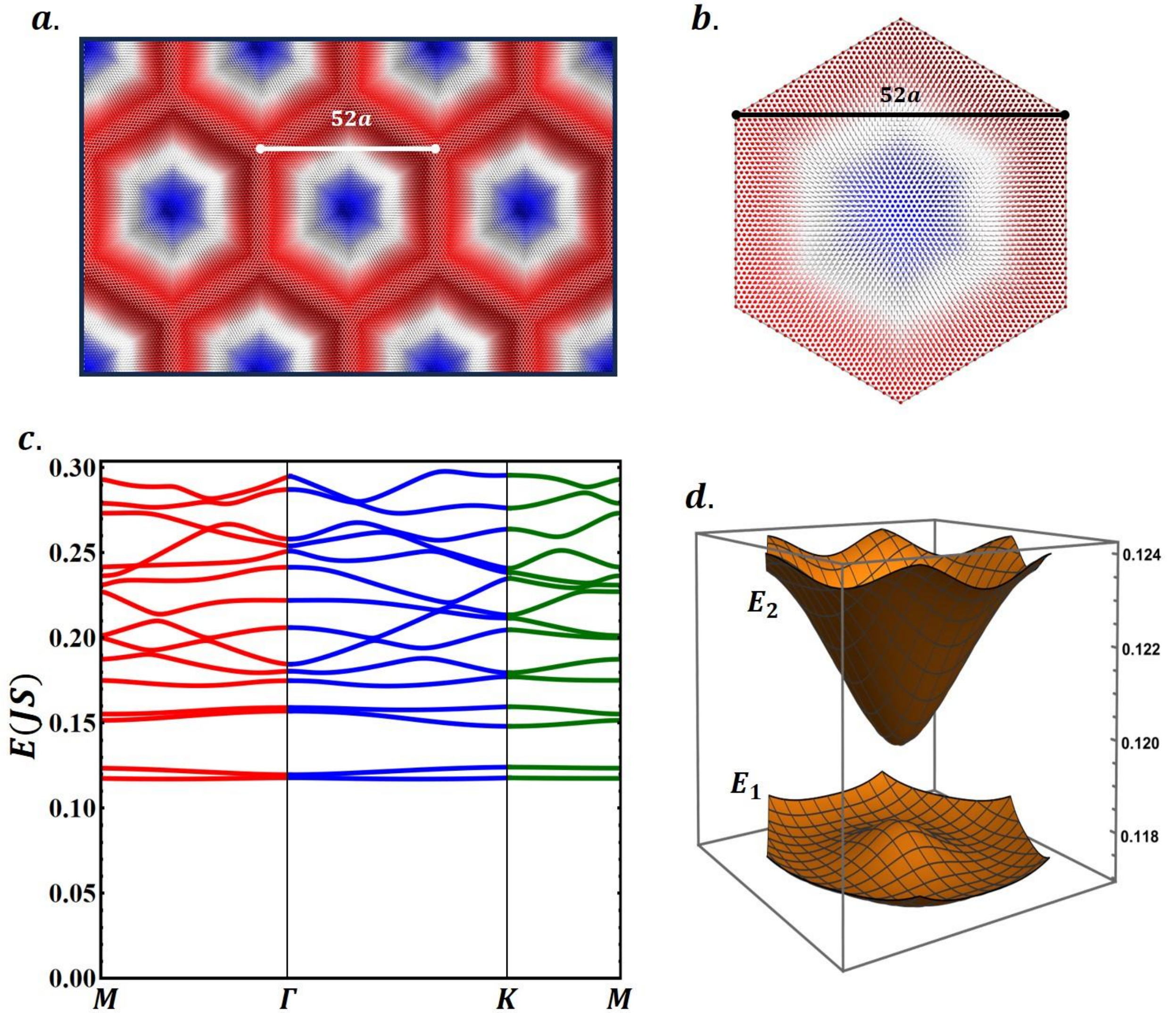

**Figure 5.** (a) The skyrmion lattice and (b) a single skyrmion comprising 3661 spins, obtained for a DMI strength of $D = 0.1\,J$ and a magnetic field of $B = 0.175\,T$. In this case, the skyrmion's width and lattice period are $52a$. Figure (c) shows the dispersion curves for the lowest 15 magnon bands corresponding to the skyrmion lattice in (a). (d) A 3D plot over the BZ showing the first ($E_1$) and the second ($E_2$) energy bands, highlighting the narrow bandwidth dispersion of the first energy band and the gap between $E_1$ and $E_2$.

The size of the skyrmion, which correlates with the strength of the magnetic field, not only influences the profile of the magnon band structure but also significantly modifies its topology. Contrary to the minimal magnetic field case, at $B = 0.175\,T$, the entire band structure becomes topological, except for the two lowest energy bands (Table 1). The Berry curvatures for the topological bands at $B = 0.175\,T$ are depicted in Figure 6. Notably, in the 15-band spectrum at $B = 0.175\ T$, the (relatively) high-energy bands ($n \geq 5$) are topological and densely packed, approaching the profile of a topological band continuum.

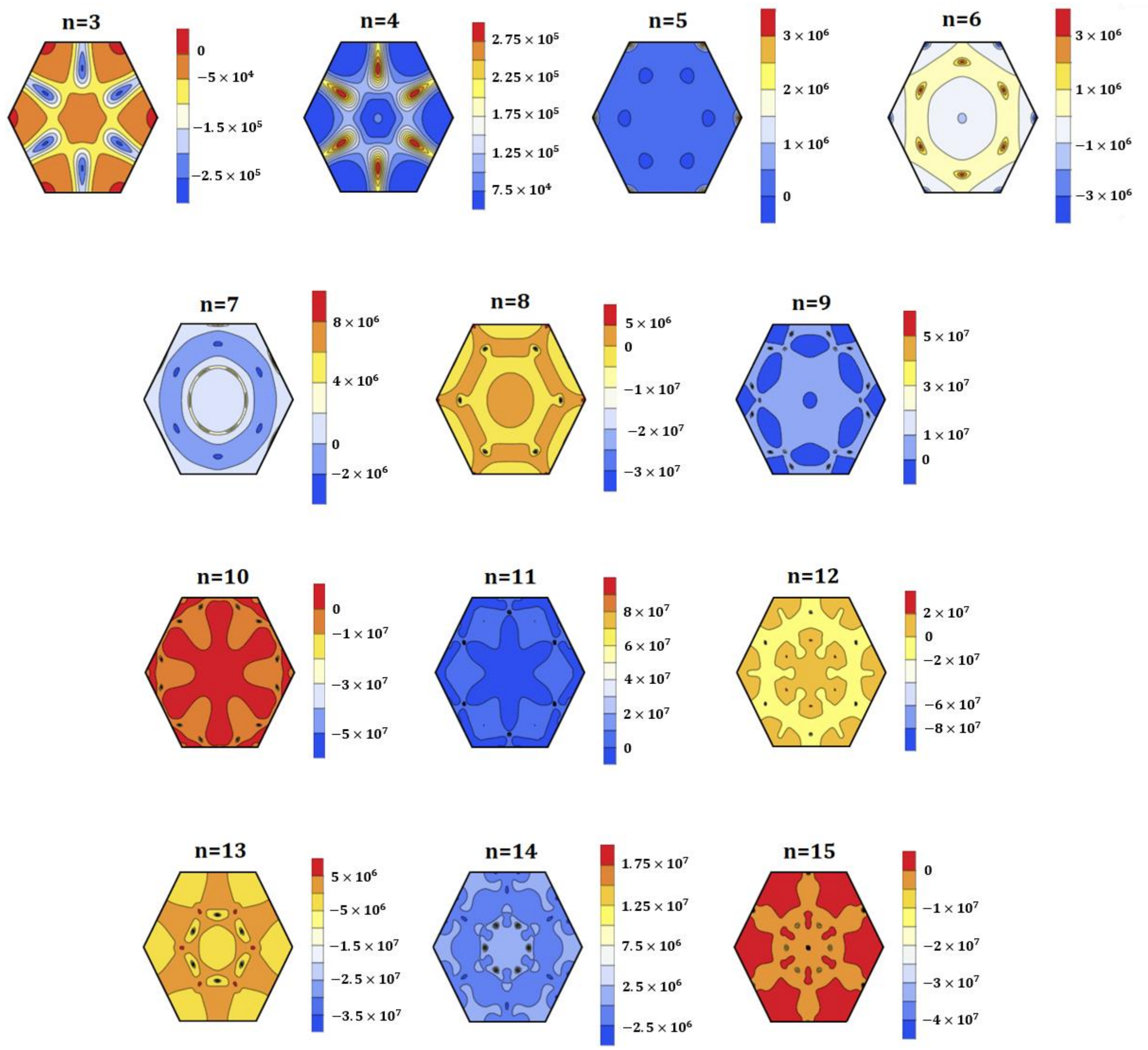

**Figure 6.** Berry curvatures for the topological magnon bands at a magnetic field of $0.175\ T$. The corresponding Chern numbers are listed in the third row of Table 1.

Our findings thus far have shown a profound dependence of the magnon band structure and its topology on the magnetic field (or skyrmion size). To further explore this dependence, we consider next the maximal magnetic field value of $0.225\ T$. At this magnetic field, the skyrmion is formed of 2755 spins (Figure 7a, b). The 15-band spectrum for this case is presented in Figure 7c, showcasing a significantly different profile compared to the previous magnetic fields. The entire band structure shifts to higher energies, increasing the band gap for the lowest energy mode, while the highest energy band reaches a value near $0.38\,J$. All bands are dispersive at the maximal magnetic field, showing that flat bands are not a universal feature of the skyrmionic magnon band structure. The entire low-energy band structure becomes densely packed, resembling a band continuum (Figure 7c). More importantly, the calculation of the Berry curvature and Chern numbers reveals that all bands are topological (Table 1), with some Chern numbers reaching significant magnitudes (e.g., $C_{13} = -6$ and $C_{14} = 8$). The Berry curvatures for all 15 bands are illustrated in Figure 8.

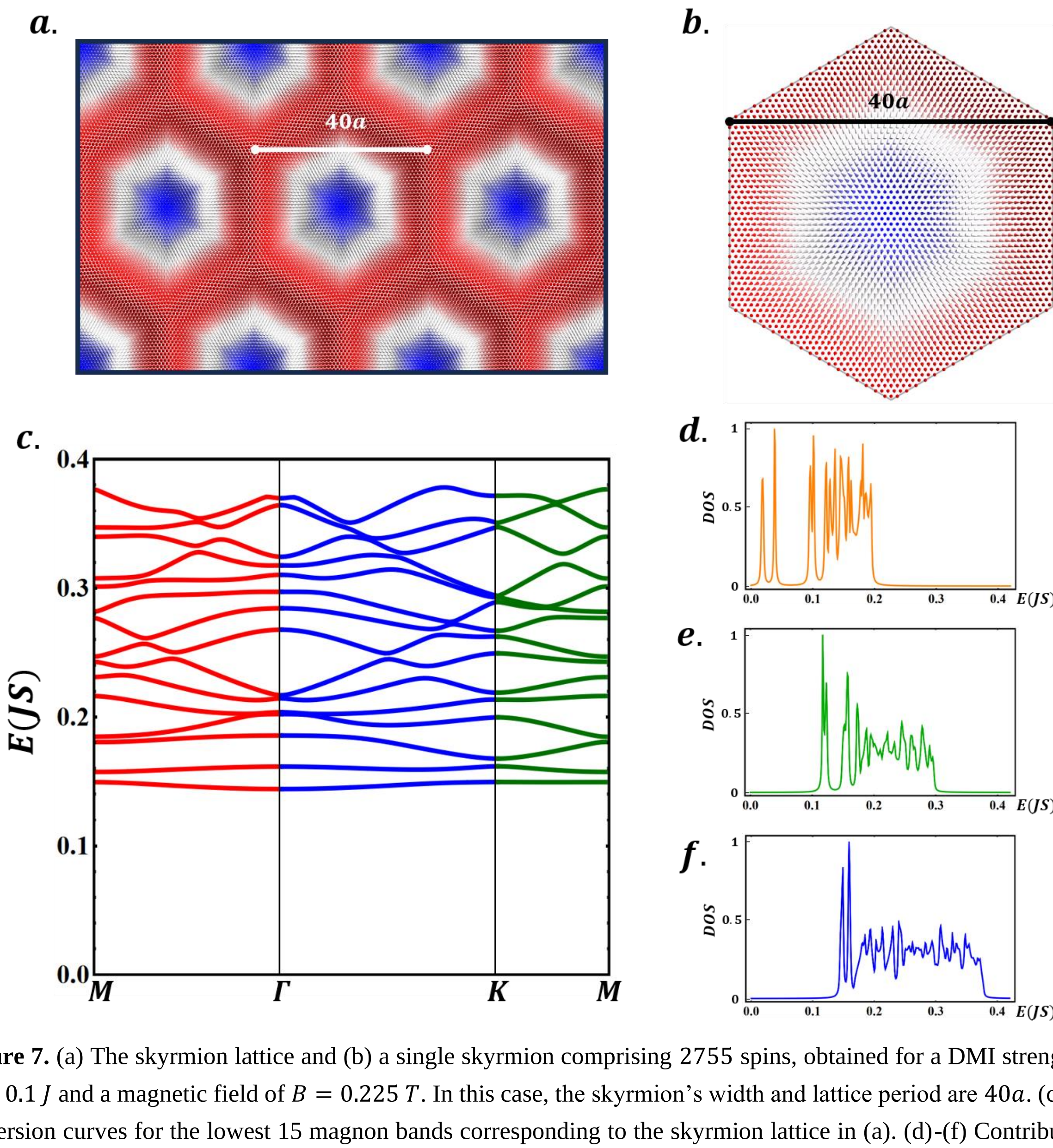

**Figure 7.** (a) The skyrmion lattice and (b) a single skyrmion comprising $2755$ spins, obtained for a DMI strength of $D = 0.1\,J$ and a magnetic field of $B = 0.225\,T$. In this case, the skyrmion's width and lattice period are $40a$. (c) The dispersion curves for the lowest 15 magnon bands corresponding to the skyrmion lattice in (a). (d)-(f) Contributions of the 15 lowest energy bands to the density of states (DOS) for magnetic fields of 0.125 T, 0.175 T, and 0.225 T, respectively. In these plots, the DOS is normalized with respect to its maximum value.

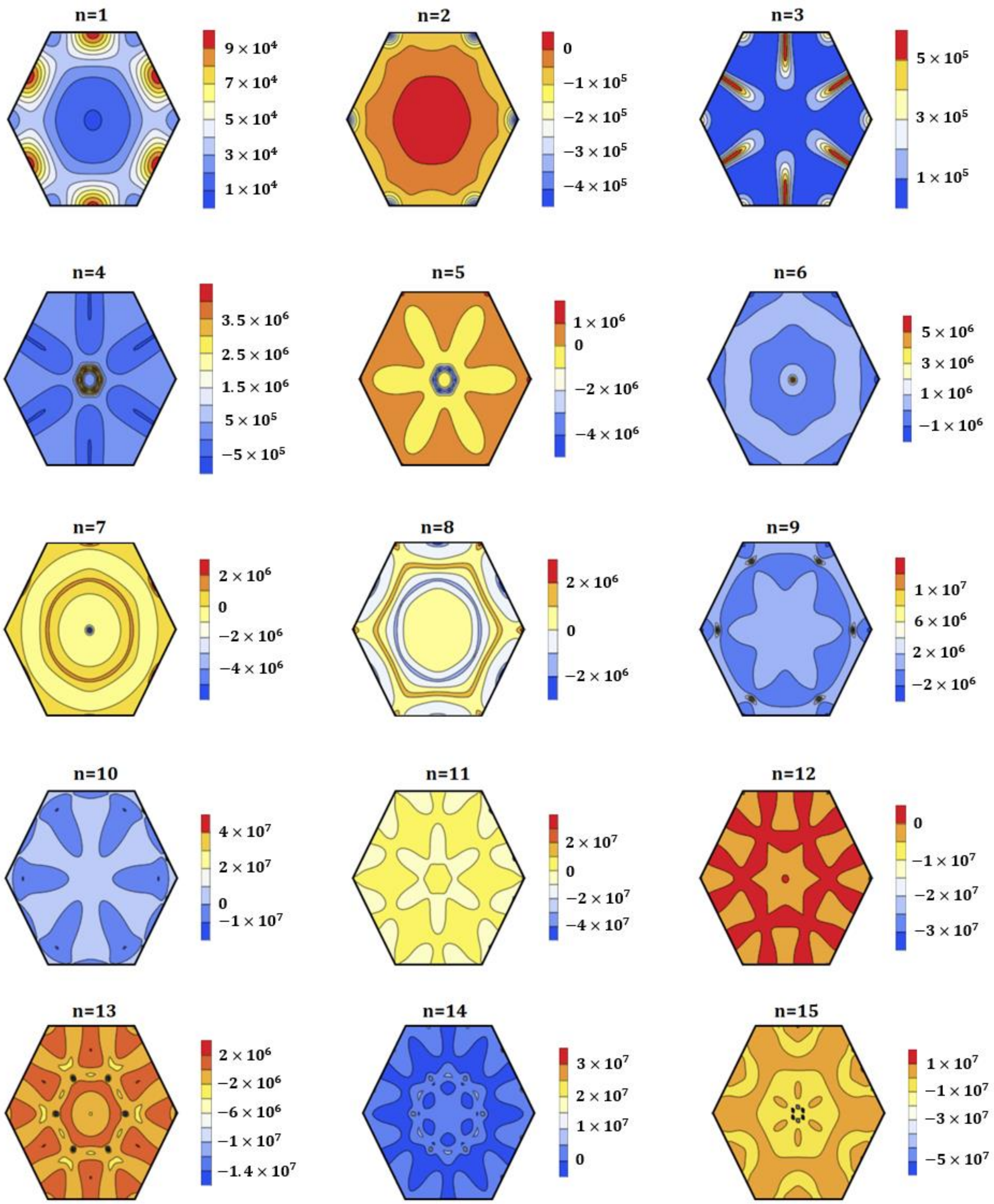

**Figure 8.** Berry curvatures for the topological magnon bands at a magnetic field of $0.225\ T$. The corresponding Chern numbers are listed in the fourth row of Table 1.

| | $C_1$ | $C_2$ | $C_3$ | $C_4$ | $C_5$ | $C_6$ | $C_7$ | $C_8$ | $C_9$ | $C_{10}$ | $C_{11}$ | $C_{12}$ | $C_{13}$ | $C_{14}$ | $C_{15}$ |
|---|---|---|---|---|---|---|---|---|---|---|---|---|---|---|---|
| $\mathbf{0.12\ T}$ | 0 | 0 | 0 | 0 | 1 | 0 | 0 | -1 | 0 | 0 | 5 | -5 | 5 | -2 | 2 |
| $\mathbf{0.175\ T}$ | 0 | 0 | -1 | 2 | 2 | 1 | 3 | -5 | 6 | -5 | 6 | -4 | -1 | -1 | -3 |
| $\mathbf{0.225\ T}$ | 1 | -1 | 3 | 1 | -1 | 2 | 2 | 1 | 1 | 3 | -1 | 1 | -6 | 8 | -4 |

**Table 1.** Chern numbers ($C_n$) for the lowest 15 magnon bands ($n = 1$ to $n = 15$, in order of increasing energy) at different magnetic fields: $0.12\ T$ (second row), $0.175\ T$ (third row), and $0.225\ T$ (fourth row).

The multiple topological phase transitions likely arise from numerous band gap closing and reopening scenarios, predominantly involving the tiny gaps observed in our spectra. Importantly, these gap closures naturally involve bands beyond the 15 lowest energy bands, particularly due to the increasing band density at higher energies. Given the intricate band structure profiles and their high sensitivity to the magnetic field, tracking the numerous gaps as they close and reopen becomes challenging. Therefore, we demonstrated the multiple topological phase transitions by calculating the Chern numbers at minimal, moderate, and maximal magnetic fields. Calculating Chern numbers is a reliable method for confirming topological phase transitions, especially since band gap closing and reopening do not guarantee the occurrence of such transitions [9].

The large Chern numbers observed in some bands at relatively high energies can be attributed to the compaction effect induced by the magnetic field. These bands are densely packed, resembling a continuum, with numerous tiny gaps between them across various regions of the BZ. This configuration generates pronounced peaks in the Berry curvatures, which can ultimately result in large Chern numbers. Additionally, as previously mentioned, the band gap closing and opening is expected to involve high-energy bands beyond the first 15 bands. The density of these high-energy bands is expected to be even greater, further promoting the emergence of large Chern numbers that might transfer to lower energy bands as gaps close and reopen.

Finally, we have calculated the contributions of the 15 lowest energy bands to the density of states, which are depicted in Figures 7d-f for the magnetic fields of 0.125 T, 0.175 T, and 0.225 T, respectively. It is observed that higher magnetic fields promote the merging of peaks in the density of states due to the compaction effect previously discussed. However, the overall density of states decreases at higher magnetic fields (see Figure S3 of the Supplementary Notes [20]), primarily due to the reduction in skyrmion size and the consequent increase in the BZ area.

## IV. Discussion

We developed a momentum-space theory to describe magnon dynamics in large 2D ferromagnetic skyrmions as an alternative to the real-space cluster approach. Using Fourier-space techniques and a continuum treatment of the skyrmions, we derived the magnon Hamiltonian exclusively in terms of momentum-space operators. In this picture, umklapp scattering of magnons becomes inevitable, leading to a multi-momentum Hamiltonian. The profile and topology of the low-energy magnon spectrum derived from this Hamiltonian are found to be critically dependent on the magnetic field, and thus the skyrmion's size, which can render the entire spectrum topological.

The momentum-space theory predicts multiple topological phase transitions induced by the external magnetic field, in contrast to the single topological phase transition revealed by the real-space cluster approach for small skyrmions [9]. In small skyrmions, the topological phase transition occurs at a specific magnetic field due to the closure of the gap between the third band (the counterclockwise mode) and the fourth band (the breathing mode) at the center of the BZ. Meanwhile, the multiple topological phase transitions predicted by the momentum-space theory can be attributed to numerous gap closing and reopening scenarios. A comprehensive study to identify these various gap closures and the modes involved is highly demanding and merits a dedicated investigation.

We anticipate that the standard skyrmionic magnon modes (e.g., breathing, clockwise, counterclockwise, etc.) predicted in previous studies [3,7–9,11,12,14–16] will also appear in the band structures of the momentum-space theory. However, rigorous investigation is required to confirm this in future studies. Indeed, determining skyrmion distortions due to specific modes will necessitate further development of the momentum-space theory, including numerically demanding inverse Fourier transformations to recover the real-space spins and obtain the skyrmion's distortion.

Theoretically, topological (bulk) bands are expected to induce topologically protected edge states according to the bulk-edge correspondence. Therefore, the rich topological band structures revealed in our calculations suggest the potential emergence of multiple topologically protected edge modes in finite samples, warranting further investigation. However, the emergence of topological edge states requires that the finite skyrmion lattice contains an integer number of unit cells [9,42]. Consequently, stripes formed from large skyrmions would have a significant width, further complicating the theoretical study.

Our study concluded that flat bands are not universal features of the magnonic spectrum. Specifically, we observed a single relatively flat band at low and moderate magnetic fields, which disappeared at a higher magnetic field. Flat bands are often associated with localized distortions of individual skyrmions in the skyrmion lattice [3,7–9,11,12,14–16], resembling the modes of isolated skyrmions. As noted in Reference [9], the overlap between neighboring skyrmions decreases as their size shrinks, promoting the isolated skyrmion limit and the flattening of some magnon bands in small skyrmions. This argument aligns with the absence of ideally flat magnon bands in the momentum-space theory for large skyrmions.

The lowest-energy band is gapped near the BZ's center, indicating the absence of the Goldstone mode for the studied skyrmions. Nevertheless, the inverse relationship observed between the lowest-energy gap and the skyrmion size suggests that this mode could potentially be retrieved for extremely large skyrmions at very low DMI strengths. However, exploring this possibility at such extreme conditions is beyond the scope of this manuscript.

We emphasize that our work specifically addresses 2D magnets without generalizing the conclusions to 3D magnets or magnetic thin films. Topological spin textures have been reported in several 2D magnets [43–55]. However, observing their magnons remains experimentally challenging, mainly due to the limitations of current measurement techniques and the difficulty in achieving sufficiently ordered skyrmion lattices [4,31–33,56]. Our focus on 2D magnets justifies the exclusion of dipolar interactions, which are generally considered weak in such materials [57–62].

The proposed theory awaits significant future developments. This includes an extension to other types of spin lattices, such as the honeycomb lattice, which characterizes several 2D magnets [55,63–65]. Additionally, it is crucial to integrate other fundamental interactions, such as intrinsic next-nearest-neighbor DMI (IDMI), Kitaev interactions, and magnetic anisotropies. Given that the IDMI and Kitaev interaction can induce 2D topological magnons even in a collinear ground state [57–62,66–70], examining their effects in skyrmion ground states offers an exciting avenue for further research. Furthermore, applying this theory to other topological spin textures [71], such as antiferromagnetic skyrmions, antiskyrmions, and bi-merons, would also be of considerable relevance.

Finally, we stress that both the real-space cluster theory and the momentum-space theory are essential for studying magnons in different classes of skyrmion lattices. The first approach is suited for the class of small or quantum skyrmions [14,15,23,72], while the second is applicable to the more common class of large or classical skyrmions. The divergence in results does not invalidate either approach, as both are valid when applied to their respective length scales.

## Acknowledgments

Part of the numerical calculations was performed using the Phoenix High Performance Computing facility at the American University of the Middle East (AUM), Kuwait.